\documentclass{article}
\usepackage{PRIMEarxiv}
\usepackage[utf8]{inputenc} % allow utf-8 input
\usepackage[T1]{fontenc}    % use 8-bit T1 fonts
\usepackage{hyperref}       % hyperlinks
\usepackage{url}            % simple URL typesetting
\usepackage{booktabs}       % professional-quality tables
\usepackage{amsfonts}       % blackboard math symbols
\usepackage{nicefrac}       % compact symbols for 1/2, etc.
\usepackage{microtype}      % microtypography
\usepackage{lipsum}
\usepackage{fancyhdr}       % header
\usepackage{graphicx}       % graphics
\graphicspath{{media/}}     % organize your images and other figures under media/ folder
\usepackage{amsfonts}
\usepackage{amssymb}
\usepackage{graphicx}% Include figure files
\usepackage{dcolumn}% Align table columns on decimal point
\usepackage{bm}% bold math
\usepackage{amsmath}
\usepackage{mathrsfs}
\usepackage{cases}
\usepackage{bm}% bold math
\usepackage{color}
\usepackage[pagewise]{lineno}% add line numbers
\usepackage{subcaption}
\usepackage{braket}
\usepackage{ragged2e}
\usepackage{hyperref}% add hypertext capabilities
\newlength{\normalitemsep}
\normalitemsep=\itemsep
\newlength{\normalparskip}
\normalparskip=\parskip
\newenvironment{parenum}{%
    \setlength{\itemsep}{\normalitemsep}% 
    \setlength{\parskip}{\normalitemsep}% 
    \begin{enumerate}
    \setlength{\itemsep}{\normalitemsep}%
    \setlength{\parskip}{\normalitemsep}%
    \setlength{\itemsep}{\normalitemsep}% 
    \setlength{\parskip}{\normalitemsep}% 
}%
{\end{enumerate}%
}

\title{Bubble $^{36}Ar$ and Its New Breathing Modes}            
 \author{
 	Ge Ren \\
 	College of Physics, Henan Normal University, China, Xinxiang \\
 	Shanghai Advanced Research Institute, Chinese Academy of Sciences, China, Shanghai \\
 	\And
 	Chun-Wang Ma $*$ \\
 	College of Physics, Henan Normal University, China, Xinxiang \\
 	Shanghai Research Center for Theoretical Nuclear Physics, NSFC and Fudan University, China, Shanghai \\
 	Institute of Nuclear Science and Technology, Henan Academy of Sciences, China, Zhengzhou \\
 	\And
 	Xi-Guang Cao $*$ \\
 	Shanghai Advanced Research Institute, Chinese Academy of Sciences, China, Shanghai \\
 	Shanghai Institute of Applied Physics, Chinese Academy of Sciences, China, Shanghai \\
 	\And
 	Yu-Gang Ma $*$ \\
 	Shanghai Research Center for Theoretical Nuclear Physics, NSFC and Fudan University, China, Shanghai \\
 	Key Laboratory of Nuclear Physics and Ion-beam Application (MOE), \\
 	Institute of Modern Physics, Fudan University, China, Shanghai \\
 }           

\begin{document}
\maketitle
\begin{abstract}
	%% Text of abstract
	The bubble nuclei are important components of exotic nuclear structures characterized by special depletions of central densities. Focusing on bubble structures of $^{36}$Ar, the characterizations of bubble nuclei were explored with the framework of the extended quantum molecular dynamics model. Three density distribution modes were uncovered for the first time, i.e. micro-bubble, bubble, and cluster resonances, which show unique spectral signature compared to the monopole resonance spectrum as the excitation intensity was increased in bubbles. Of pivotal importance is the revelation that the bubble mode's oscillation frequency closely resembles macroscopic bubble dynamics, building a connection between classical macroscopic phenomena and the quantum complexity of the nuclear structure. The discovery marks a crucial step forward in deciphering the relationship between classical and quantum domains within the enigmatic world of atomic nuclei.
\end{abstract}

\renewcommand{\thefootnote}{ }
\footnote{$*$ Co corresponding author: machunwang@126.com (Chun-Wang Ma), caoxg@sari.ac.cn (Xi-Guang Cao), mayugang@fudan.edu.cn (Yu-Gang Ma)}
\section{Introduction}
\label{introduction}

As the emergence of large-scale radioactive ion beam facilities, the exploration of exotic nuclei continues to reshape nuclear physics \cite{01.TANIHATA_PLB1992307_exotic_Li11,02.KORSHENINNIKOV_NPA2005501_exotic_experimental}, enhancing our understanding of atomic nuclei and the quantum mechanism behind \cite{03.Otsuka_PhysRevLett.97.162501_Exotic_Shell.Structure,04.Gaudefroy_PhysRevLett.109.202503_Light.Exotic,05.YANG_PPNP2023104005_exotic_spectroscopy}. One of the most intriguing frontiers in contemporary nuclear physics research is the exotic nuclear structure. Despite the relative comprehensiveness of traditional and structural models in explaining conventional nuclei, they face limitations when tackling singular nuclei. In recent years, significant strides have been made in studying exotic nuclei, ranging from calculations involving various deformed nuclei \cite{06.JoliePhysRevLett.89.182502_deformed_Triple.Point,07.Gaffney_Nature2013_deformed_pear-shaped,08.Schenke_nst_2024_hexadecapole_deformation} to investigate unique cluster configurations such as linear chains \cite{09.Morinaga_PhysRev.101.254_Linear-Chain,10.Liu_PhysRevLett.124.192501_Linear-Chain,11.Han_Commun.Phys.2023_Linear-Chain}, toroidal structures \cite{12.Ichikawa_PhysRevLett.109.232503_Torus,13.Cao_PhysRevC.99.014606_ring,Liao2024} and other states \cite{13.5.Kawabata_nst2024_5a,13.8.ma2024-0013}. Of special interest is the critical category of bubble nuclei, a concept initially posited by Wilson in the 1940s, which describes a depletion of central density in nuclei similar to the soap ``bubble'' \cite{14.Wilson_PhysRev.69.538_Spherical.Shell,15.Wilson_PhysRev.77.516_Spherical.Shell}. The phenomenon originates from the extremely low $s_{1/2}$ state occupying probability with the absence of $l=$ 0 contribution \cite{16.KHAN_NPA200837_bubble,17.GRASSO_IJMPE200910.1142_BUBBLE}. This unusual density distribution raised fundamental questions about its origins, implications, and, notably, the collective behavior of excited nuclei during heavy-ion collisions.

Since 1970s, Wong has discussed toroidal nuclei and spherical bubble in liquid-drop model \cite{18.WONG_PLB1972446_Toroidal,19.WONG_AP1973279_Toroidal.bubble}. Campi and Sprung conducted their first theoretical calculations on $^{36}$Ar and $^{200}$Hg \cite{20.CAMPI_PLB1973291_bubble}. In subsequent decades, the possibility of the existence of bubble structures has been explored in light \cite{21.Ebran_Nature2012_bubble.cluster}, medium \cite{22.Grasso_PhysRevC.79.034318_bubble,23.Wang_PhysRevC.84.044333_bubble,24.Choudhary_PhysRevC.102.034619_bubble}, heavy, and super-heavy nuclei \cite{25.Bender_PhysRevC.60.034304_Shell.bubble,26.DECHARGE_NPA200355_Bubbles.semi-bubbles}. Mutschler \textit{et al.} marched experimentally toward the bubble structure of $^{34}$Si by measuring the $\gamma$-ray spectrum of the one-proton removal reaction in 2017 \cite{27.Mutschler_NaturePhysics2017_bubble}. 

Although these studies described well the density distribution of bubble nuclei using nuclear structure models, the description of their dynamic behavior after excitation in reactions still lacks. Due to the difficulty in forming fully enclosed bubble structures in light-mass nuclei with limited nucleons and the challenge to create almost completely hollow bubbles in heavy-mass nuclei with a large number of nucleons, the research targets at the in-medium mass region for exploration. Inspired by the pioneering work on bubble $^{36}$Ar by Campi and Sprung \cite{20.CAMPI_PLB1973291_bubble}, and given the significant interest in its potential cluster structure, research focus has shifted towards understanding the excited states in its bubble structure. Of particular note is the role of the isoscalar monopole resonance colloquially known as the ``breathing mode'' of collective motion in nuclei, which refers to the vibrational excitation mode of nuclei while maintaining their spherical symmetry. It is conceivable that the density distributions of nuclei are liquid droplets or bubbles have a different overall impact on the vibration in this kind of spherically symmetric vibrational mode. It is specifically aimed to investigate how the unique ``bubbles” central structure of certain nuclei influences their monopole resonance behavior under differing levels of excitation intensity, and to identify possible new breathing modes.

\section{Methodology}
\label{Analyse}

The extended quantum molecular dynamics (EQMD) model refined by Maruyama \textit{et al}., which is widely recognized for its efficacy in investigating cluster structures \cite{28.He_PhysRevLett.113.032506_EQMD,29.Dey_PhysRevC.102.031301_EQMD}, was employed to investigate the monopole resonance of $^{36}$Ar in bubble structure. The main advantage of the EQMD model includes the subtraction of the zero-point kinetic energy, the integration of an effective Pauli potential into the interaction mechanics, and the incorporation of a complex variable wave packet width into the Hamiltonian \cite{30.Maruyama_PhysRevC.53.297_EQMD}. These collective enhancements enable the EQMD model to initialize systems that more closely approximate the ground state and the bound state of excited nuclei, which facilitates the research on the properties of bubble structures within them. 

After obtaining the initial state of a nucleus system $\Psi(0)$ at $t=0$ fm/\emph{c} in the EQMD model, $\Psi(0)$ could be boosted to excited systems, 
\begin{eqnarray}\label{eq1.psi.M.psi}
        \bra{\Psi^{'}} = \mathcal{\hat{M}} \ket{\Psi (0) },
\end{eqnarray}
where $\mathcal{\hat{M}}$ is the perturbation excitation operator. Because that the system contains $A$ nucleons, an operator on it can be seen as the sum of all single particle operators $\mathscr{\hat{M}}_{i}$, and the expectation of $\mathcal{\hat{M}}$ is,
\begin{eqnarray}\label{eq2.M^}
\begin{aligned}
<\mathcal{\hat{M}}> 
& = \bra{\Psi^{'}} \mathcal{\hat{M}} \ket{\Psi^{'}} = \sum \limits_{i}^{A} \bra{\Psi^{'}} \mathscr{\hat{M}}_{i} \ket{\Psi^{'}}\\
& = \int d \vec{r}_i d {\vec{r}_i}^{~'} \rho ({\vec{r}_i}^{~'}, \vec{r}_i) \bra{\vec{r}_i} \mathscr{\hat{M}}_{i} \ket{{\vec{r}_i}^{~'}}.
\end{aligned}
\end{eqnarray}
Substituting the Wigner transform, one has
\begin{eqnarray}\label{eq3.Wigner.M^}
<\mathcal{\hat{M}}> = \int d \vec{r} \int d \vec{p} f(\vec{r}, \vec{p}) \mathscr{M} (\vec{r}, \vec{p}).
\end{eqnarray}
During excitations, the total Hamiltonian of the system can be decomposed into two parts in the following Hamiltonian, i.e., the unperturbed eigenstate $\hat{H}^{0}(t)$ and the applied excitation perturbation $\hat{H^{'}}(t)$,
\begin{eqnarray}\label{eq4.Hamilton}
\hat{H}(t)= \hat{H}^{0}(t) + \hat{H^{'}} (t),
\end{eqnarray}
where the perturbation $\hat{H^{'}} (t) = \varepsilon \mathcal{\hat{M}}\delta(t)$, and  $\varepsilon$ determines the magnitude of the intensity of the perturbation. A larger $\varepsilon$ represents an increased initial perturbation and indicates a greater energy input during the onset of excitation, thus serves as a measurement of the initial excitation intensity.

Because the isotropy property of the monopole resonance, the perturbation effect on the initial coordinates and momentum of particles can be expressed as,
\begin{eqnarray}\label{eq5.rp}
    \left\{
    \begin{aligned}
        {\vec{r}_i}^{~'} &= \vec{r}_{i}+\varepsilon\frac{\partial \mathscr{M}(\vec{r}_{i},\vec{p}_{i})}{\partial\vec{p}_{i}},\\
        {\vec{p}_i}^{~'} &= \vec{p}_{i}-\varepsilon\frac{\partial \mathscr{M}(\vec{r}_{i},\vec{p}_{i})}{\partial\vec{r}_{i}},
    \end{aligned}
    \right.
    \implies
    \left\{
    \begin{aligned}
        {\vec{r}_i}^{~'} &= \vec{r}_i + \varepsilon \vec{r}_i,\\
        {\vec{p}_i}^{~'} &= \vec{p}_i - \varepsilon \vec{p}_i,
    \end{aligned}
    \right.
\end{eqnarray}
\section{Results and Discussion}
\label{sec.rst.disc}
\subsection{\textbf{The microscopic bubble}}
\subsubsection{Density distribution}

The wave function of the $i$-th nucleon in EQMD is
\begin{equation} \label{eq:wave_function}
\begin{aligned}
\phi_i(\vec{r})=\left(\frac{v_i+v_i^*}{2 \pi}\right)^{3/4} \exp \left[\frac{v_i}{2}\left(\vec{R}_i-\vec{r}\right)^2+\frac{i}{\hbar} \vec{P}_i \cdot \vec{r}\right],
\end{aligned}
\end{equation}
where $v_i=\frac{1}{\lambda_i}+i \delta_i$. $\lambda_i$ represents the real part of the wave packet, and $\delta_i$ symbolizes the imaginary part. $(\vec{r}, \vec{p}) $ indicates a point in the phase space, whereas $\left(\vec{R}_i, \vec{P}_i\right)$ denotes the centroid of the wave packet of the $i$-th nucleon in phase space.

Performing the Wigner transformation on Eq. (\ref{eq:wave_function}), one obtains the phase space distribution function 
\begin{equation} \label{eq:Phase_space_function}
\begin{aligned}
 f(\vec{r},\vec{p})& = \frac{1}{(2 \pi \hbar)^{3/4}} \int \exp \left(\frac{i \vec{p} \xi}{\hbar}\right) \phi_i\left(\vec{r}^{-}\right) \phi_i^*\left(\vec{r}^{+}\right) d \xi \\
& =\frac{1}{(\pi \hbar)^3} \exp \left[-\frac{1+\lambda_i^2 \delta_i^2}{\lambda_i}\left(\vec{r}-\vec{R}_i\right)^2\right] \\
& \times \exp \left[-\frac{\lambda_i}{\hbar^2}\left(\vec{p}-\vec{P}_i\right)^2\right]  \exp \left[\frac{2 \lambda_i \delta_i}{\hbar}\left(\vec{r}-\vec{R}_i\right)\left(\vec{p}-\vec{P}_i\right)\right] .
\end{aligned}
\end{equation}
in which $\vec{r}^{\pm}=\vec{r} \pm \frac{\xi}{2}$ is defined, and \(\xi \in \mathbb{R}\) is an auxiliary variable used in the construction of the integral.
The density distribution function of nucleus is the result of integrating $f(\vec{r}, \vec{p})$ with respect to $\vec{p}$
\begin{equation}
\begin{aligned}
 \rho_i(\vec{r})&=\int f(\vec{r}, \vec{p}) d \vec{p}, \\ 
 \rho(\vec{r})&=\sum_i \rho_i(\vec{r}) =\sum_i \frac{1}{\left(\pi \lambda_i\right)^{3 / 2}} \exp \left[-\frac{\left(\vec{r}-\vec{R}_i\right)^2}{\lambda_i}\right] .
\end{aligned}
\end{equation}

\begin{figure}
    \centering
    \includegraphics[width=0.5\columnwidth]{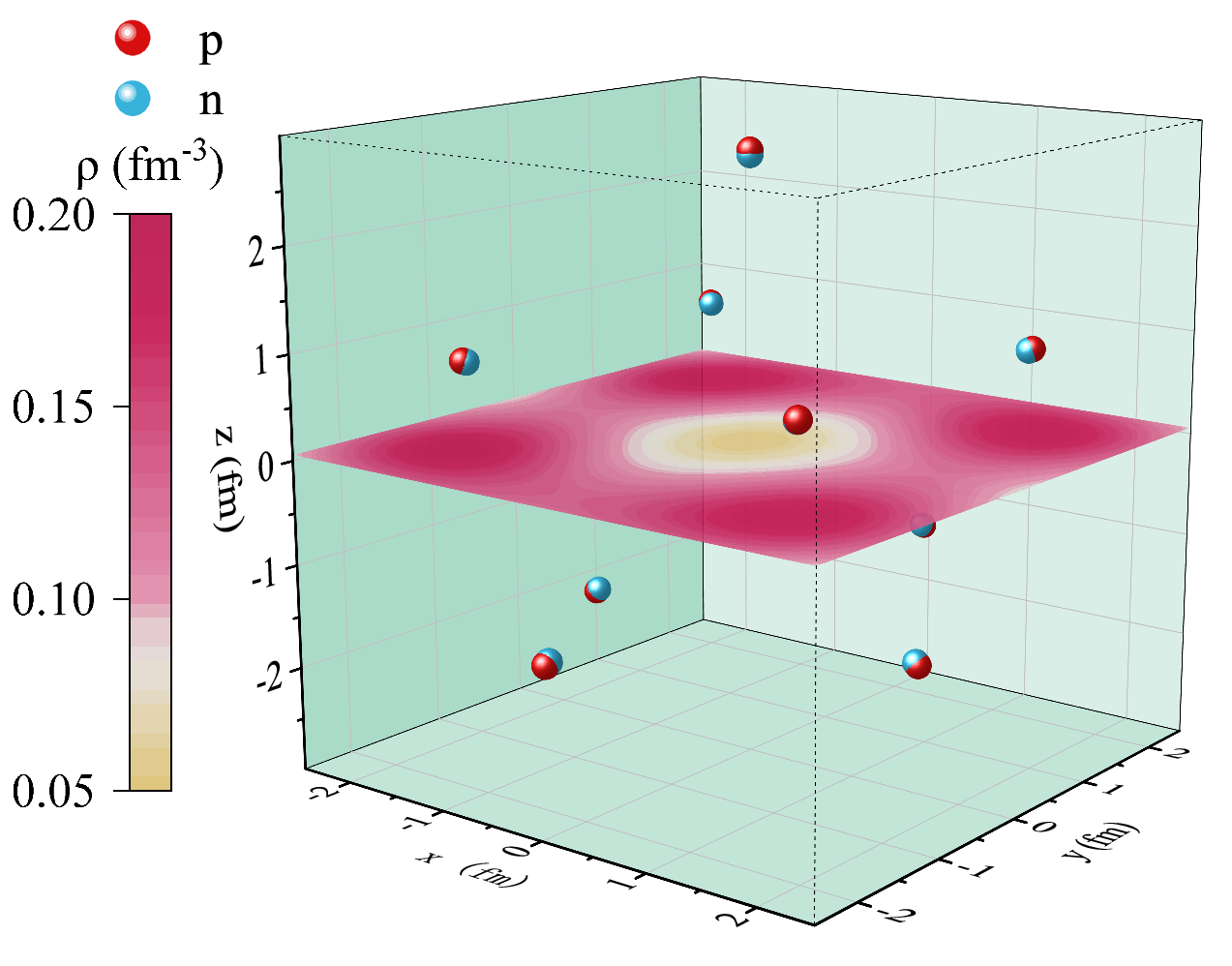}
    \caption{The 9$\alpha$ clusters of the ground state of $^{36}$Ar, obtained through the initialization process prior to excitation in the EQMD simulation, and their corresponding density distribution of the $z=0$ profile.}
    \label{fig0.9a}
\end{figure}

The ground state of $^{36}$Ar can be well obtained through the initialization process prior to excitation in the EQMD simulation, which reveals spheroidal-like configuration of 9$\alpha$ clusters. A typical configuration of clustered $^{36}$Ar obtained from simulation and the density distribution of its $z=0$ profile are depicted in Fig.~\ref{fig0.9a}. The relative difference between the calculated binding energy of 8.146 MeV/u and the experimental value of 8.520 MeV/u is 4.388 \%, while the relative difference between the calculated charge radius of 3.0539 fm and the experimental value of 3.3905 fm is 9.93 \%. 

\begin{figure}[!h]
 \centering
\begin{subfigure}[b]{0.4\columnwidth}
    \centering
    \includegraphics[width=\columnwidth]{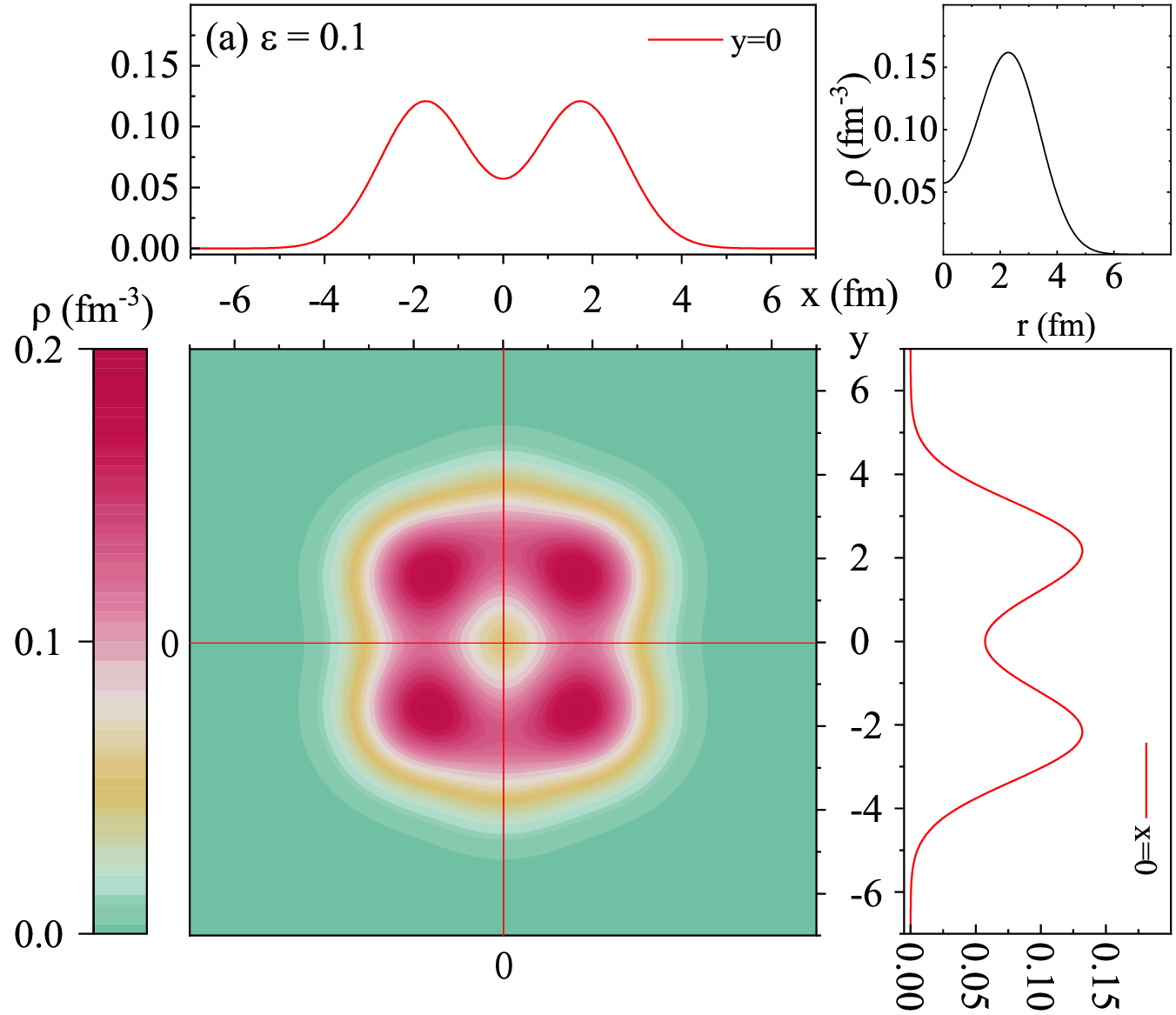}
    % \caption{}
    \label{fig.1_0.1}
\end{subfigure}
% \par\bigskip %
\begin{subfigure}[b]{0.4\columnwidth}
    \centering
    \includegraphics[width=\columnwidth]{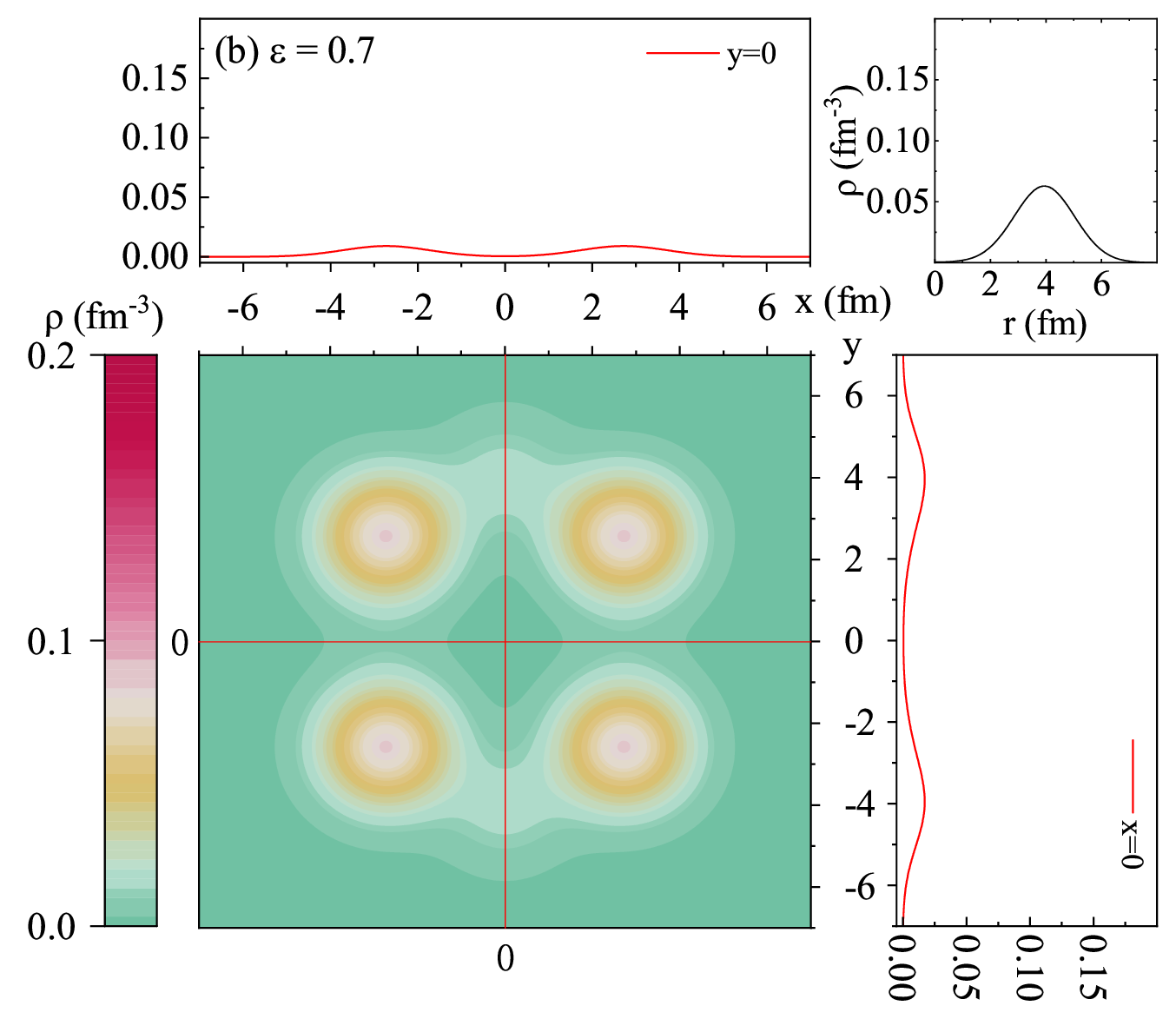}
    % \caption{}
    \label{fig.1_0.7}
\end{subfigure}
% \par\bigskip %
\begin{subfigure}[b]{0.4\columnwidth}
    \centering
    \includegraphics[width=\columnwidth]{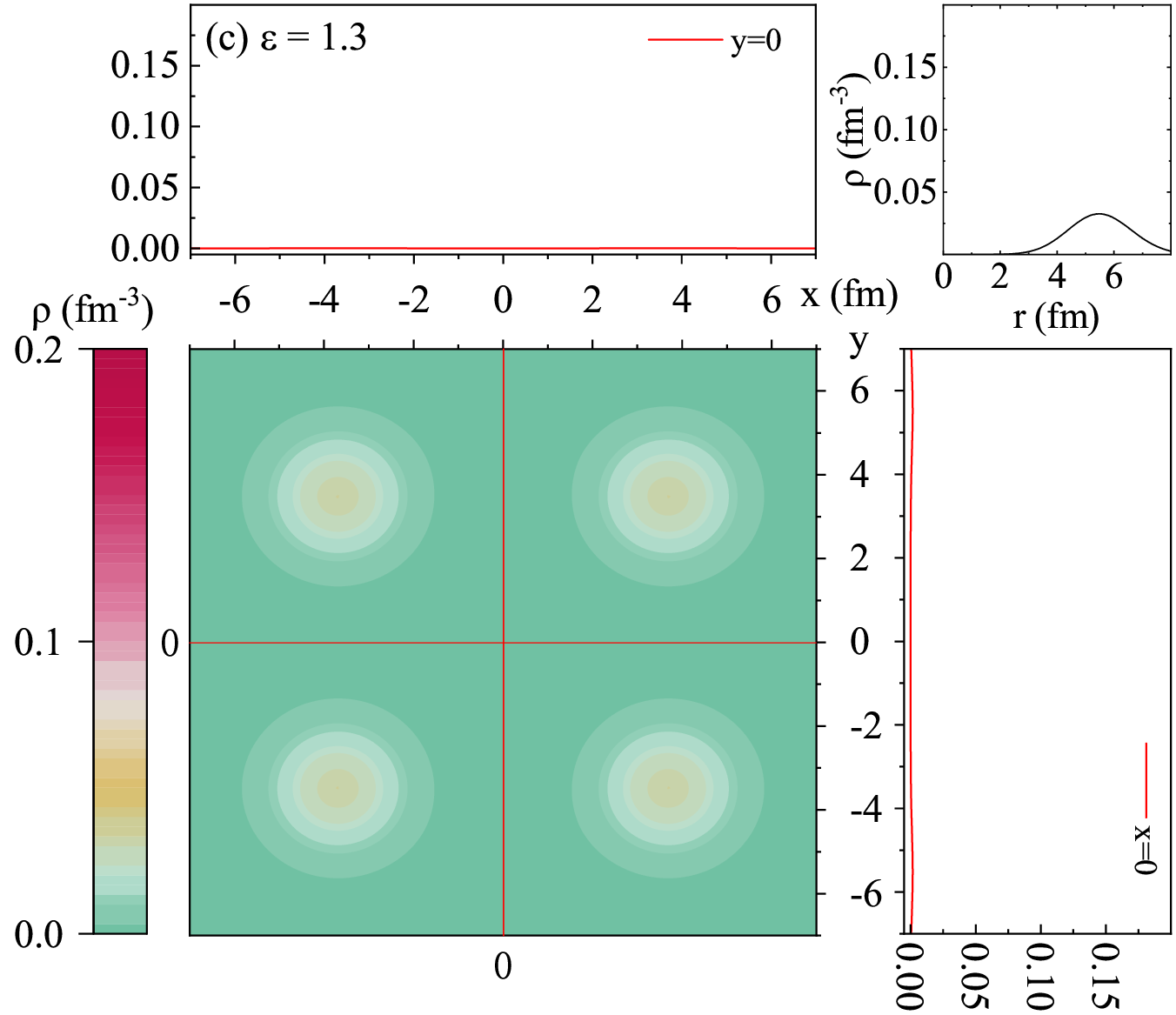}
    % \caption{}
    \label{fig.1_1.3}
\end{subfigure}
\caption{\justifying\label{fig.1_density}The density distribution (left lower) and average radial density (upper right) of the $^{36}$Ar nuclear profile at $\varepsilon = 0.1, 0.7, 1.3$.}
\end{figure}

In such a cluster-composed nucleus, the density distribution can be assumed as a distinctive profile, resembling a central ``bubble'' encircled by a denser nuclear hull or liquid film. To facilitate the visualization and analysis of the density distribution, the centre of mass is designated as the origin of the coordinate axis, and the nucleus is suitably rotated to ensure a symmetric distribution of the nuclear wave packet's centre relative to this axis. With an increasing value of $\varepsilon$, the density distribution exhibits three different situations. Figure \ref{fig.1_density} illustrates these density distributions in the $\emph{z=0}$ profile for each scenario. It appears to consist of four $\alpha$ clusters because the four nearest $\alpha$ clusters above the plane exhibit a more pronounced density distribution on it. The curves positioned in the upper left and lower right corners of the figure, respectively, depict the density distributions along the $\emph{y}=$ 0 and $\emph{x}=$ 0 axes, respectively. Meanwhile, the average radial density distribution is shown in the upper right corner. Three different modes are illustrated;
\begin{parenum}
\item The microbubble structure mode. For $\varepsilon=$ 0.1, a distinct low-density centre surrounded by dense clusters is shown, with the outer layer exhibiting a thicker high-density region, resembling a small liquid bubble characterized by a low-density centre surrounded by a thick surface;
\item The bubble mode. For $\varepsilon=$ 0.7, despite a visible separation between the clusters, a thin film persists in connecting them. Concurrently, the density in central area nearly drops to 0, akin to a soap bubble;
\item The cluster resonance mode. For $\varepsilon=$ 1.3, the thin films that connect the clusters are completely detached, resulting in independent density distributions.
\end{parenum}

Ebran \textit{et al.} \cite{21.Ebran_Nature2012_bubble.cluster} studied the impact of cluster distances on nuclear states, and postulated a classification scheme dividing them into three principal categories, i.e., the quantum liquids, clusters, and crystal-like structures. Importantly, these three classifications align with and are supported by findings in this study of bubble $^{36}$Ar nucleus. 

\subsubsection{Root mean square radius}

\begin{figure}[htbp]
\centering
\includegraphics[width=0.5\columnwidth]{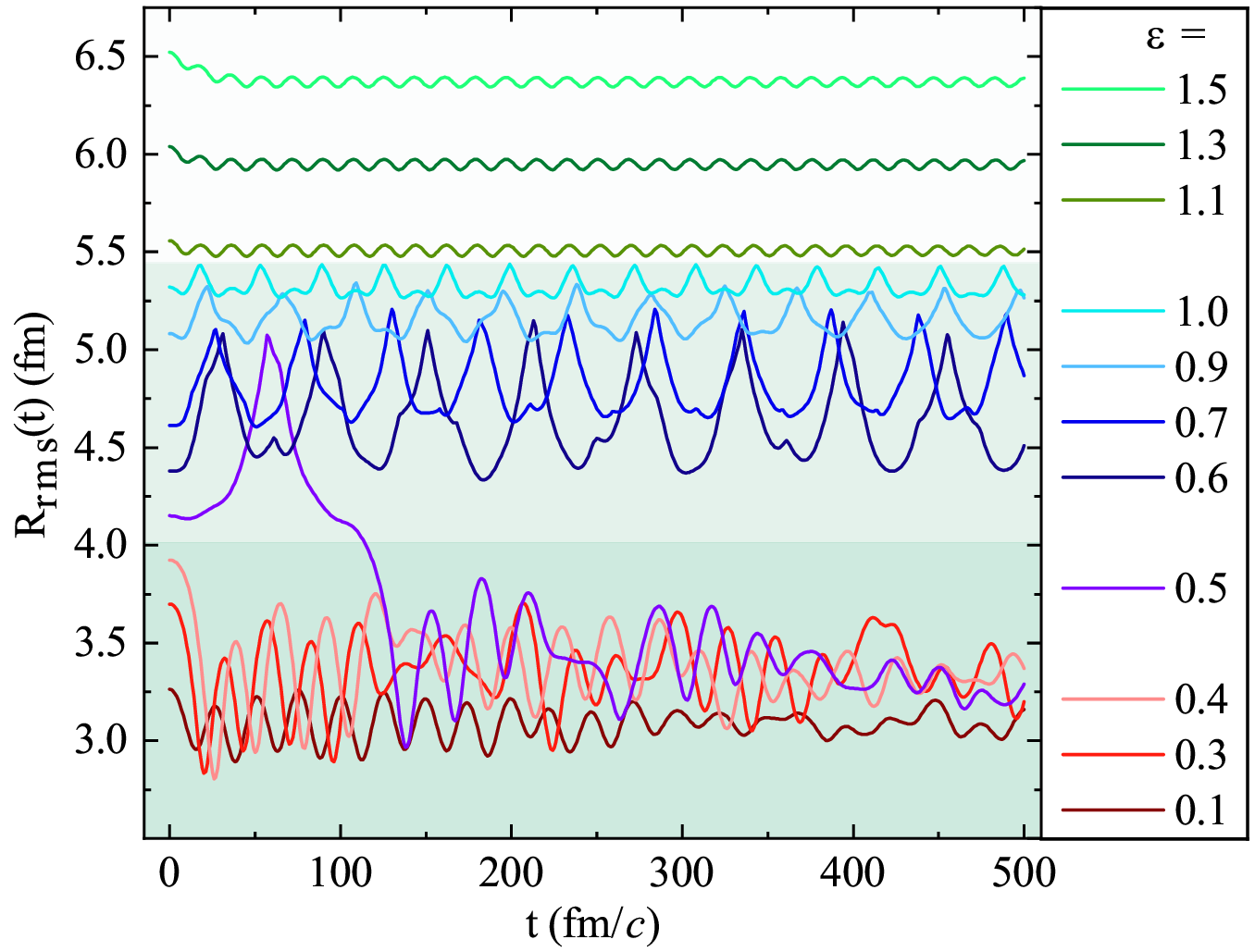}
\caption{\justifying\label{fig2_RMS} The evolution of the root-mean-square radius of $^{36}$Ar nuclei simulated by the EQMD model for $t=$ 0 - 500 fm/\emph{c} with $\varepsilon$ from 1 to 1.5. The shadows are used to distinguish results of three modes.
}
\end{figure}

During the radial expansion and compression phases of the monopole resonance, the radius of the mean square root ($R_{rms} (t)$) of the nucleus is used as the moment of the monopole $R_{rms} (t)$ for each time step \cite{31.BARAN_NPA2001373_rms,32.Gaitanos_PhysRevC.81.054316_rms} (see the lines in Fig.~\ref{fig2_RMS}). With increasing $\varepsilon$, $R_{rms} (t)$ displays three distinct oscillation patterns on $t$, i.e., $0 <\varepsilon< 0.5$, $0.5 <\varepsilon < 1.0$, and $1.0 < \varepsilon < 1.5$, as distinguished by shadows at different depths in Fig. \ref{fig2_RMS}. Furthermore, $\varepsilon = 0.5$ and $\varepsilon = 1.0$ mark the critical values between these forms. The following conclusions can be reached, 
\begin{parenum}
\item Within $0 <\varepsilon< 0.5$, corresponding to the microbubble structure mode, $R_{rms} (t)$ shows an oscillatory structure characteristic of traditional giant monopole resonance. Owing to the mean-field potential and dissipative effects of nucleon-nucleon scattering, the oscillation remains relatively stable initially but starts to diminish after 150 - 200 fm/\emph{c}, followed by rapid damping after 300 fm/\emph{c};
\item Within $0.5 <\varepsilon < 1.0$, corresponding to the bubble mode, the oscillation of $R_{rms}(t)$ shows more consistent variation over time, with each cycle featuring a sharp peak followed by a smooth valley; 
\item Within $1.0 < \varepsilon < 1.5$, corresponding to the cluster resonance mode, the oscillation of $R_{rms} (t)$ remains nearly unchanged over time, showing an extremely small amplitude.
\end{parenum}

It should be noted that, at $\varepsilon =$ 0.5, the bubble $^{36}$Ar nucleus was shortly excited to the second mode within the initial 100 fm/\emph{c}. Nevertheless, the insufficient excitation intensity failed to sustain the vibrational of the second mode, leading to its eventual collapse to the first mode.

Given the relatively steady changes in $R_{rms} (t)$ within each vibrational mode, they have been verified to correspond to the three modes of density distribution. The oscillatory behavior of the $R_{rms} (t)$ in microbubbles bears resemblance to conventional monopole resonances, with a marginally extended periodicity. In contrast, the resonance phenomena observed between bubbles and cluster configurations diverge from traditional patterns, exemplifying the characteristics of new breathing modes.

\subsubsection{\texorpdfstring{$\gamma$}{}-ray spectra}
Since impact of density distribution and shape on the excitation energy, the $\gamma-$ray spectra are commonly used as a means of studying nuclear cluster resonance. The vibrational frequency $f_n$ of $R_{rms} (t)$ and the $\gamma-$ray spectra can be obtained by performing the Fourier transform after taking the second-order derivative of time. And the probability spectrum of photon energy $E_{\gamma} = \hbar \omega$ is represented by, which could serve as an experimental observation \cite{28.He_PhysRevLett.113.032506_EQMD},
\begin{eqnarray}
    \label{eq6.rms}
    R_{rms}^{\prime\prime}(\omega)=\int_{t_{0}}^{t_{max}}R_{rms}^{\prime\prime} (t)e^{i\omega t}dt , 
\end{eqnarray}        
\begin{eqnarray} 
    \label{eq7.P/E}
    \frac{d \mathcal{P}}{dE_\gamma}=\frac{2e^2}{3\pi\hbar c^3E_\gamma}|R_{rms}^{\prime\prime}(\omega)|^2,
\end{eqnarray}
where ${d \mathcal{P}}/{dE_\gamma}$ can be interpreted as the nuclear photon absorption cross section. 

\begin{figure}[h!]
\centering
\includegraphics[width=0.5\columnwidth]{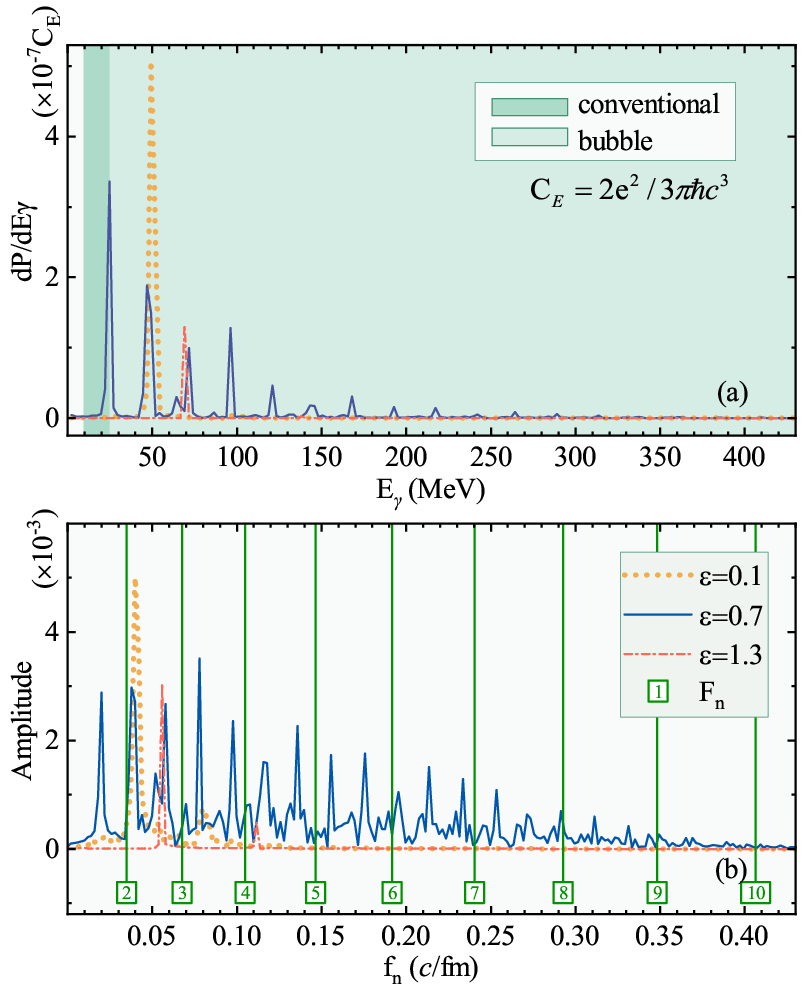}
\caption{\justifying{\label{fig3_Egamma_frequency}  (a) The $\gamma$-ray spectra of $^{36}$Ar at $\varepsilon = $ 0.1, 0.7, and 1.3, respectively. In Eq.~\ref{eq7.P/E}, we define the constant term $C_{E}=2e^{2} / 3\pi\hbar c^{3}$. The shadows are used to distinguish the peak range of conventional monopole resonance and the bubble modes. (b) The monopole resonance frequency $f_n$ and the corresponding relative intensities. By substituting the surface tension coefficient $\sigma$ of nuclear matter into Eq.~(\ref{eq10.Fn}), the frequency peaks of $F_n$  (green open squares) can be obtained using the method of macroscopic bubbles, where the numbers represent the mode $n$.}
}
\end{figure}

In Fig.~\ref{fig3_Egamma_frequency}(a), the probability spectrum of the photon energy $E_\gamma$ is plotted. The conventional monopole resonance $\gamma$ emission spectra of nuclei typically exhibit a prominent peak within the energy range from 10 to 25 MeV. In contrast, the new breathing mode characteristic of the bubble nuclei deviates from this norm. The microbubble and cluster resonance modes exhibit a prominent peak at about 50 and 70 MeV, whereas the bubble mode displays multiple peaks with decreasing intensity as the energy increases. These peaks in the $\gamma$ spectra hold experimental probe to these isoscalar monopole resonances.

\subsection{\textbf{Bridging the microscopic and macroscopic bubbles}}

\subsubsection{Frequency}

The vibrational frequencies $f_n$ of $R_{rms} (t)$ are plotted in Fig.~\ref{fig3_Egamma_frequency}~(b), which exhibit similar characteristics to the $\gamma$ ray spectra. Interestingly, Horace successfully described the macroscopic bubble vibrations in the fluid mechanics \cite{33.HORACE_soapbubble_lamb1879_soap}. For a macroscopic bubble, the eigenmode of the system is defined by the spherical harmonic $Y_{nm}$. Given the axial symmetry of the oscillation, $m = 0$, and its characteristics are solely determined by the mode number $n$. Taking the origin at the centre, the shape of the common surface at any instant is given by \cite{33.HORACE_soapbubble_lamb1879_soap,34.Rayleigh_PLMSs1-10_soap}, 
\begin{eqnarray}
    \label{eq9.R}
    R=R_{0}+Y_{n}\sin(\omega t+\varepsilon),
\end{eqnarray}
where $R_0$ is equilibrium radius. The natural frequency of the eigenmode during its free vibration correlates with the equilibrium radius $R_0$, surface tension coefficient $\sigma$, fluid density $\rho$, and mode number $n$ \cite{33.HORACE_soapbubble_lamb1879_soap,34.Rayleigh_PLMSs1-10_soap,35.Lundgren_Mansour_JFM1988_soap,36.Kornek_NJP2010_soap},
\begin{eqnarray}
    \label{eq10.Fn}
    F_{n}=\frac{1}{2\pi}\sqrt{\frac{\sigma}{\rho {R_{0}}^{3}}(n-1)n(n+2)},
\end{eqnarray}
where $n = 0, 1, 2,\dots,$. According to the experimental findings measured by Kornek \textit{et al.}, \cite{36.Kornek_NJP2010_soap}, the intensity of vibration decreases with the mode number $n$, which suggests the potential correlation between the  found bubble patterns in bubble $^{36}$Ar nucleus and the experimental macroscopic bubble system. 

\subsubsection{Pressure}
\label{sec_Pressure}
For a macroscopic bubble in the equilibrium state, the pressure difference between the internal ($P_{in}$) and external ($P_{ex}$) is described by the Young-Laplace formula \cite{37.YoungAnEO_PTRS_press,38.P.S.Laplace_press},
\begin{eqnarray}
    \label{eq11.Y_L_Pressio}
    P_{in}-P_{ex}=\frac{2\sigma}{R}.
\end{eqnarray}
In thermodynamics and statistical physics, the pressure is calculated on the basis of the momentum. The nucleus was split along the radial direction, with each small segment represented as $\delta r$. Given that the mean radial density peak possesses internal and external half-heights denoted by $R_{in}$ and $R_{ex}$ respectively, the momentum of $\delta r$ at $R_{in}$ in the direction of motion during breathing can be expressed as,
\begin{equation}\label{eq12.dp}
\begin{aligned}
d p_{R_{in}} & =d m_{\delta r} \cdot v_{R_{in}}  =d \rho_{R_{in}} \cdot d V_{\delta r} \cdot v_{R_{in}} \\
& =d \rho_{R_{in}} \cdot 4 \pi R_{in}^2 \cdot \delta r \cdot \frac{\partial R_{in}}{\partial t}. 
\end{aligned}
\end{equation}
According to the relationship between impulse and momentum, the instantaneous pressure acting on $R_{in}$ is
\begin{equation}\label{eq13.dFR1}
\begin{aligned}
d P_{R_{in}} =\frac{d F_{R_{in}}}{d S_{\delta r}} =\frac{2 d p_{R_{in}}}{4 \pi R_{in}^2 d t} =\frac{2 \delta r d \rho_{R_{in}} \frac{\partial R_{in}}{\partial t}}{d t}. 
\end{aligned}
\end{equation}

Due to the structural and vibrational similarities between bubble nuclei and macroscopic bubbles, we assume that the bubble nuclei satisfy the Young-Laplace formula at the equilibrium position of vibration, similar to macroscopic bubbles \cite{33.HORACE_soapbubble_lamb1879_soap}. By substituting the pressures on both sides of the bubble nuclei liquid film into Eq.~(\ref{eq11.Y_L_Pressio}), the surface tension coefficient $\sigma$ can be determined for the bubble mode of $^{36}$Ar across a range of excitation levels with different $\varepsilon$. Substituting $\sigma$ into Eq.~(\ref{eq10.Fn}), the vibrational frequency $F_n$ can be obtained using the method of macroscopic bubbles and the peak position is shown in Fig.~\ref{fig3_Egamma_frequency}(b) with labeled numbers. Subsequently, a correlation between the natural frequencies of the observed  macroscopic bubble vibrational modes and the predicted microscopic vibrational modes can be established.

\begin{figure}[h!]
\centering
\includegraphics[width=0.75\columnwidth]{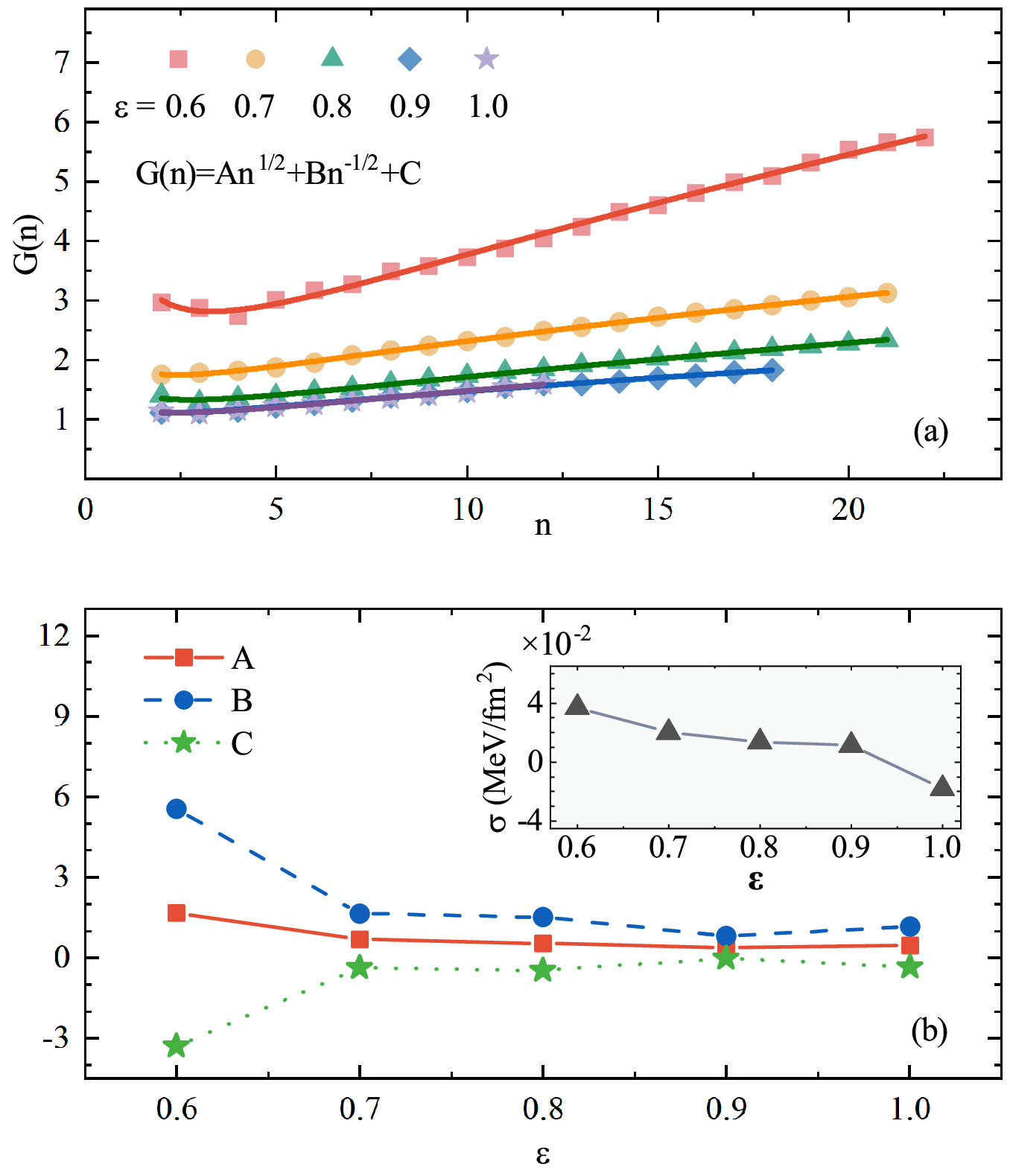}
\caption{\justifying{\label{fig4._Gn}  (a) The value of $F_n/f_n$ changing with mode number $n$. The line represents the correlation function $G(n)$ derived through fitting over the range of $\varepsilon =$ 0.6 to 1.0. (b) The parameters $A, B$ and $C$ of the correlation function $G(n)$ and the surface tension coefficient $\sigma$ (inset) changing with $\varepsilon$.}
}
\end{figure}

\subsubsection{Correlation function}

Due to the influence of the liquid film density distribution on the spherical harmonic at the bubble nucleus surface, it is therefore necessary to transform Eq.~(\ref{eq9.R}) to,
\begin{eqnarray}\label{eq15.R_EQMD}    
R=R_{0}+\mathcal{\hat{G}}_{n}Y_{n}\sin(\omega t+\varepsilon),
\end{eqnarray}
where $\mathcal{\hat{G}}_{n}$ denotes the operator changing in the density distribution of mode $n$. Equation~(\ref{eq10.Fn}) is transformed to 
\begin{eqnarray}\label{eq16.fn_EQMD}    
    f_{n}=\frac{1}{2\pi G(n)}\sqrt{\frac{\sigma}{\rho R_{0}^{3}}(n-1)n(n+2)}.
\end{eqnarray}
$G(n)$ is the function of mode $n$,
\begin{eqnarray}\label{eq17.Gn}
    G(n)=An^{\frac{1}{2}}+Bn^{-\frac{1}{2}}+C,
\end{eqnarray}
which obeys the hook function as denoted by fitting lines in Fig.~\ref{fig4._Gn}(a).
The fitted values of $A$, $B$, and $C$ all approach 0 with increasing $\varepsilon$, while the variation of them with $\varepsilon$ is depicted as seen in Fig.~\ref{fig4._Gn}(b).

The increase of excitation intensity $\varepsilon$ makes the decreases of both the thickness and density of the bubble's liquid film, alongside a decrease in the surface tension coefficient $\sigma$ [see Fig.~\ref{fig4._Gn}(b)]. This trend can be attributed to the diminishing nuclear force among $\alpha$ clusters as their separation increases, resulting in a reduction in the surface tension coefficient. At $\varepsilon =$ 1.0, $\sigma < 0$ marks the boundary between the bubble mode and the cluster resonance mode. This consists to the phenomenon of macroscopic scenario where external pressure exceeding internal pressure prompts the bubble to rupture.

For atomic nuclei, the frequency at which mode $n$ is larger can be attributed to the ripple-like effects caused by the vibrations of $\alpha$ clusters within the nuclear structure. This results in a greater difference between the density of nuclei and the macroscopic bubble concepts at positions that are closer and farther from the $\alpha$ cluster. Consequently, the correlation function $G(n)$ exhibits a characteristic hook shape. As $\varepsilon$ increases, the distances between the $\alpha$ clusters grow larger, which in turn diminishes their relative spatial occupancy on the surface of the bubble. Therefore, the overall structure of the nuclear bubble becomes more akin to a macroscopic bubble, with the frequency ratio between them being closer to 1, and parameters $A$, $B$, and $C$ in $G(n)$ tending to 0.

Different to studies on nuclear bubble structures in neutron-rich and proton-rich regions, this study focuses on neutron-proton symmetric $^{36}$Ar, which illustrates a bubble structure characterized by $\alpha$ clusters. It is quite different to the proton bubbles due to the relocated proton density distribution in neutron-rich nuclei. The bubble $^{36}$Ar structures indicate more work worth exploring in the exotic excited states in nuclei near the $\beta$-stable line. Besides, for the increasing interest on superheavy nuclei synthesis with nuclear fusion, the thin nuclear film in bubble nuclei of in-medium nucleus may shed new light for its reduced fusion potential barrier between the target nucleus. 

\section{Conclusions}
To conclude, the monopole resonance modes within the bubble nucleus $^{36}$Ar is investigated in the framework of the EQMD model. Three distinct new breathing modes are revealed, i.e., the microbubble, the bubble, and the cluster resonance modes, with the increased excitation intensity. The remarkable correlation between the vibrational frequency of the bubble mode and the macroscopic concept of bubble is illustrated, which bridges the gap between the microscopic and macroscopic aspects of bubble phenomena and uncovers intriguing parallels between the minima and maxima in physics across both scales. The findings in this letter extend beyond the boundary of bubble nuclei, providing insights into the broader field of nuclear physics, as well as to nuclear structure and the response of nucleons under extreme conditions.

\section*{Acknowledgements}
CXG thanks the Strategic Priority Research Program of Chinese Academy of Sciences (No. XDB34030000), the National Key Research and Development Program of China (No. 2022YFA1602404), the National Natural Science Foundation of China (NNSFC) (No. U1832129). MCW thanks the NNSFC (No. 12375123), and Natural Science Foundation of Henan Province (No. 242300421048). MYG thanks NNSFC (contract No. 12147101), the Guangdong Major Project of Basic and Applied Basic Research (No. 2020B0301030008), and the STCSM (Grant No. 23590780100).

\nocite{*}
\bibliographystyle{unsrt} 
\bibliography{Bubble_36Ar_arxiv}

\typeout{get arXiv to do 4 passes: Label(s) may have changed. Rerun}
\end{document}